\documentclass[prl,twocolumn,showpacs,nofootinbib,preprintnumbers,amsmath,amssymb,aps,longbibliography,superscriptaddress,floatfix]{revtex4-2}
\usepackage{graphicx,subfigure,epsfig}
\usepackage{dcolumn}
\usepackage{amssymb}
\usepackage{times}
\usepackage{amsmath}
\usepackage{amsfonts}
\usepackage{mathrsfs}
\usepackage{setspace}
\usepackage{latexsym}
\usepackage{bbm}
\usepackage{float}
\usepackage{flafter}
\usepackage{bm}
\usepackage{epstopdf}
\usepackage{color}
\usepackage{multirow}
\usepackage{physics} 
\usepackage[export]{adjustbox}
\DeclareMathOperator{\sgn}{sgn}
\usepackage{braket}

\usepackage{tikz}                  

\usepackage{dsfont}

\usepackage{soul} 

\def \be {\begin{eqnarray}}
\def \ee {\end{eqnarray}}

\newcommand{\bk}{{\bf k}}

\newcommand{\bq}{{\bf q}}

\newcommand{\br}{{\bf r}}

\newcommand{\bv}{{\bf v}}

\usepackage{dsfont}

\usepackage{xcolor}

 \usepackage{simpler-wick} 

\newcommand{\bea}{\begin{equation} \begin{aligned}}
\newcommand{\eea}{\end{aligned} \end{equation} }

\newcommand{\bpm}{\begin{pmatrix}}
\newcommand{\epm}{\end{pmatrix}}

\newcommand{\cc}[1]{\langle{#1}\rangle_{\rm c}}

\usepackage{xr}
\usepackage[pdftex,breaklinks,bookmarks=false,colorlinks,linkcolor=blue,citecolor=blue,urlcolor=blue,hypertexnames=false]{hyperref}

\usepackage{orcidlink}

\begin{document}

\title{Singular three-point density correlations in two-dimensional Fermi liquids}

\author{Pok Man Tam \orcidlink{0000-0001-9620-4689}}
\affiliation{Princeton Center for Theoretical Science, Princeton University, Princeton, NJ 08544}

\author{Charles L. Kane \orcidlink{0000-0002-9551-2177}}
\affiliation{Department of Physics and Astronomy, University of Pennsylvania, Philadelphia PA 19104}

\date{\today}

\begin{abstract}
We characterize a singularity in the equal-time three-point density correlations that is generic to two-dimensional interacting Fermi liquids.
In momentum space where the three-point correlation is determined by two wavevectors $\bq_1$ and $\bq_2$, the singularity takes the form $\abs{\bq_1\times\bq_2}$. 
We explain how this singularity is sharply defined in a long-wavelength collinear limit.
For a non-interacting Fermi gas, the coefficient of this singularity is given by the quantized Euler characteristic of the Fermi sea, and it implies a long-range real space correlation favoring collinear configurations.
We show that this singularity persists in interacting Fermi liquids, and express the renormalization of the coefficient of singularity in terms of Landau parameters, for both spinless and spinful Fermi liquids. Implications for quantum gas experiments are discussed.
\end{abstract}

\maketitle

\noindent {\color{blue}\emph{Introduction--}} 
Universal features of gapless quantum phases of matter are often encoded in the singular behavior of response and correlation functions \cite{abrikosovmethods, mahan2000many, Giuliani_Vignale_2005, Chubukov2003}.   For a phase described at low energy by a conformal field theory (CFT), the long distance real space correlations, which reflect the universal CFT data, are associated with non-analytic singularities in the momentum space correlations \cite{BigYellowBook, Fradkin2013book}.  For systems without conformal symmetry, including Fermi systems that exhibit a Fermi surface, much less is understood.  
Recently, we have established that for a two-dimensional non-interacting Fermi gas, the Euler characteristic $\chi_F$ of the Fermi sea is encoded in the three-point correlations of the density, according to
\cite{TamKane2022a, TamKane2023b}
\begin{equation}\label{eq: topological formula, momentum space}
    s_3(\bq_1, \bq_2) = \frac{\abs{\bq_1 \times \bq_2}}{(2\pi)^2}\chi_F,
\end{equation}
where $s_3(\bq_1, \bq_2) \equiv \int\frac{d^2\bq}{(2\pi)^2} \cc{\rho_{\bq_1} \rho_{\bq_2} \rho_{\bq}}$ 
is the equal-time three-point connected density correlation, with the fermion density $\rho_\bq = \int \frac{d^2 \bk}{(2\pi)^2}c^\dagger_\bk c_{\bk+\bq}$.
Eq.~\eqref{eq: topological formula, momentum space} is exact for a single band model of non-interacting fermions, provided ${\bf q}_{1,2}$ are small enough (in a way specified precisely below).   Multi-band effects modify $s_3$, but the corrections are analytic functions of ${\bf q}_{1,2}$, and do not modify the small ${\bf q}_{1,2}$ singularity.
In this work, we address the role of short-range electron-electron interactions.  We show that in a Fermi liquid $s_3$ is still singular for small ${\bf q}_{1,2}$, but the coefficient of the singularity is no longer quantized.       
This is reminiscent of the well understood effect of interactions on a single-channel Luttinger liquid \cite{Giamarchi2004, gogolin2004bosonization}, where the two-point density correlations exhibit a $|q|$ singularity proportional to the Luttinger parameter $K$:  $s_2(q) = K |q|/(2\pi)$, which is only quantized in the absence of interactions ($K=1$).

The $|{\bf q}_1 \times {\bf q}_2|$ form in Eq.~\eqref{eq: topological formula, momentum space} is unique in that there is a singular $|\varphi|$ dependence on the angle $\varphi$ between ${\bf q}_1$ and ${\bf q}_2$, even when $|{\bf q}_{1,2,3}| \ne 0$, where $\bq_3\equiv -\bq_1-\bq_2$. This reflects long ranged correlations in real space, when the positions $\br_{1,2,3}$ are collinear.   
This is a special property of a Fermi surface, which features low energy excitations that propagate in straight lines.  We will analyze this singularity in detail and show that despite the fact that Eq.~\eqref{eq: topological formula, momentum space} breaks down in a neighborhood around $\varphi=0$, the discontinuity in the slope of $s_3$ as a function of $\varphi$ is well defined in the limit $|{\bf q}_a|\rightarrow 0$.    This motivates us to introduce
the long-wavelength collinear (LWC) limit formulated in Eq.~\eqref{eq: q-collinear limit}. 
In this limit $\chi_F$ in Eq.~\eqref{eq: topological formula, momentum space} can  be reinterpreted as a sum over Fermi surface critical points, where the velocity ${\bf v} \perp {\bf q}_a$ \cite{Kane2022a}.  We then develop the interacting Fermi liquid theory first for spinless fermions, and then generalize to spinful fermions, showing that the singularity remains but with its coefficient modified by a dimensionless interaction parameter associated with the Fermi surface critical points. For an isotropic Fermi liquid, the renormalization of $s_3$ can be expressed in terms of the Landau parameters $F_0^s$ and $F_0^a$.   Finally, we will discuss the experimental implications for the measurement of equal-time multi-point correlations of Fermi gases using quantum gas microscopy \cite{Cheuk2015,Edge2015,Haller2015, Omran2015, Parsons2015, Cheuk2016b, Bakr_review, deJongh2025,daix2025observing, daix2025probing}.

\noindent {\color{blue}\emph{Singularity in Fermi gases--}} Recently, with experimental collaborators investigating quantum gases,  we have 
computed $s_3$ exactly for the free Fermi gas \cite{daix2025probing}.  For a circular Fermi surface of radius $k_F$, the
validity condition for Eq.~\eqref{eq: topological formula, momentum space} is
\begin{equation}\label{eq: topological validity condition}
    R_{\{\bq_a\}} \equiv \frac{\abs{\bq_1}\abs{\bq_2}\abs{\bq_3}}{2\abs{\bq_1 \times\bq_2}} \leq k_F, 
\end{equation}
where $R_{\{\bq_a\}}$ is the radius of the circle circumscribing the triangle formed by $\bq_{a=1,2,3}$. 
For general $\bq_a$, we found
\begin{equation}\label{eq: general formula for cFS}
\begin{split}
    &s_3 =\begin{cases}
        \frac{1}{(2\pi)^2}\abs{\bq_1 \times \bq_2}&\text{if } R_{\{\bq_a\}} \leq k_F, \\
        \frac{k_F^2}{4\pi} \left[1+\sum_{a=1}^3 \sigma_{a} G(\frac{\abs{\bq_a}}{2k_F})\right]&\text{if } R_{\{\bq_a\}} \geq k_F,  
    \end{cases}
\end{split}
\end{equation}
where $G(X) = \frac{2}{\pi}\left( \cos^{-1}X -X\sqrt{1-X^2} \right)\theta(1-X)$, $\theta(X)$ is the Heaviside step function and $\sigma_{a} \equiv \text{sgn}(\bq_b\cdot \bq_c)$ for $a\neq b\neq c=1,2,3$. The above result holds for a convex electron-like Fermi surface (with $\chi_F=1)$, while there is an overall minus sign for a concave hole-like Fermi surface (with $\chi_F=-1$). Eq.~\eqref{eq: general formula for cFS}
suggests a smoothening of the $\abs{\varphi}$ singularity for $\varphi\rightarrow 0$, since $R_{\{\bq_a\}}$ diverges for $\abs{\bq_1\times\bq_2}\rightarrow 0$, violating
Eq.~\eqref{eq: topological validity condition}. Nevertheless, since $R_{\{\bq_a\}}\sim\mathcal{O}(\abs{\bq})$, the sharpness of the singularity is restored in the long-wavelength limit
$\abs{\bq_a}/k_F \rightarrow 0$. 
Next, we show that this singular behavior holds for arbitrarily shaped Fermi surfaces, and is associated to certain critical points on Fermi surfaces.

\noindent {\color{blue}\emph{Long-wavelength collinear limit and Fermi surface critical points--}} 
We now suppose $\{\bq_1, \bq_2, \bq_3\}$ are small compared to $k_F$ and are nearly parallel or anti-parallel, so that they form a skinny obtuse triangle as depicted in Fig.~\ref{fig: Fig1}(a). Without loss of generality, we take $\bq_\text{max} = \bq_1$ as having the largest magnitude among $\bq_{1,2,3}$ and express 
$q_{\perp}\equiv\bq_2\cdot(\hat{\bq}_1\times\hat{z})$.
The LWC limit amounts to
\begin{equation}\label{eq: q-collinear limit}
    \abs{q_{\perp}} \ll \abs{\bq_{1,2,3}} \ll k_F, 
\end{equation}
under which Eq.~\eqref{eq: general formula for cFS} reduces to 
\begin{subequations}\label{eq: collinear formula for cFS}
\begin{align}
    s_3(\{\bq_a\}) &= \frac{\abs{\bq_1\times \bq_2}}{(2\pi)^2}\mathcal{G}\left(k_F^{-1}R_{\{\bq_a\}}\right),\\
    \mathcal{G}(X) &=1+\frac{1}{2X}(X-1)^2\theta(X-1).
\end{align}
\end{subequations}
Since $R_{\{\bq_a\}}= |\bq_2|||\bq_3|/(2 \abs{q_\perp})$, Eq.~\eqref{eq: topological formula, momentum space} applies for $\abs{q_\perp} \geq  |\bq_2||\bq_3|/(2k_F)$, while otherwise
$s_3(q_\perp) \approx \abs{\bq_1}\abs{\bq_2}\abs{\bq_3}/(16\pi^2 k_F)+q_\perp^2\abs{\bq_1}k_F/(4\pi^2\abs{\bq_2}\abs{\bq_3})$.
Thus for a fixed $\abs{\bq_a}$, $s_3$ features a $\abs{q_{\perp}}$ singularity that
is smoothened for $\abs{q_{\perp}}\lesssim \mathcal{O}(\abs{\bq_a}^2/k_F)$, as illustrated in Fig.~\ref{fig: Fig1}(b).  The $\abs{q_\perp}$ singularity thus becomes sharper as $\abs{\bq_a}/k_F\rightarrow 0$. \\

\begin{figure}
    \centering
    \resizebox{\columnwidth}{!}{\includegraphics[]{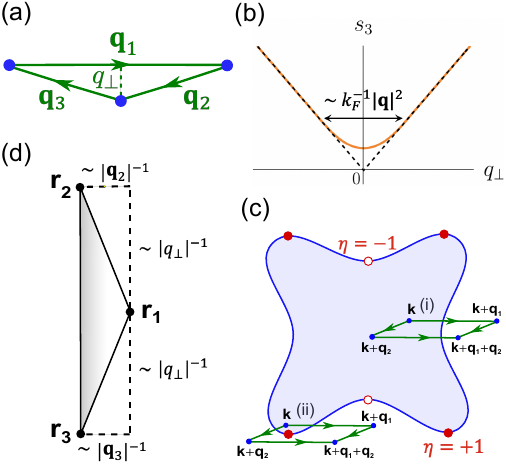}}
    \caption{(a) Momentum space triangle formed by $\{\bq_{1,2,3}\}$. (b) $s_3$ as a function of $q_{\perp}$ in the LWC limit based on Eq.~\eqref{eq: collinear formula for cFS}. The $\abs{q_{\perp}}$ singularity in 
    Eq.~\eqref{eq: topological formula, momentum space}
    (dashed line) is smoothened only for $\abs{q_{\perp}}\lesssim \mathcal{O}(\abs{\bq}^2/k_F)$. (c) In the LWC limit, $s_3$ is dominated by region near the Fermi surface critical points where $\bv \perp \bq_{a}$. The integrand $\mathcal{S}(\bk; \bq_1, \bq_2)=0$ and $1$, respectively, for situation in (i) and (ii). 
    (d) Long-range collinear correlation in real space, which corresponds to the momentum space triangle in (a). The collinear direction in real space is orthogonal to the collinear direction in momentum space.}
    \label{fig: Fig1}
\end{figure}

The LWC limit allows for a general result with an arbitrarily shaped Fermi surface. Using Wick's theorem, we can express $s_3=\int \frac{d^2\bk}{(2\pi)^2}\frac{1}{2}[\mathcal{S}(\bk; \bq_1, \bq_2)+\mathcal{S}(\bk; -\bq_1, -\bq_2)]$, with $\mathcal{S}(\bk;\bq_1,\bq_2)\equiv f_\bk(1-f_{\bk+\bq_1+\bq_2})(1-f_{\bk+\bq_1}-f_{\bk+\bq_2})$ in terms of the zero-temperature Fermi functions,
which is dominated by regions near \textit{critical points} $\bk_p$ on the Fermi surface where the velocity $\bv_p$ is perpendicular to $\bq_1$. When $\bk$ is far (compared to the scale of $\abs{\bq_a}$) from $\bk_p$, $\mathcal{S}(\bk;\bq_1,\bq_2)=0$ (Fig.~\ref{fig: Fig1}(c)(i)).  For $\bk$ near a convex critical point $\bk_p$, $\mathcal{S}(\bk; \bq_1, \bq_2)=1$ (Fig.~\ref{fig: Fig1}(c)(ii)).  For a concave critical point, $\mathcal{S}=-1$.
Therefore, we can extend Eq.~\eqref{eq: collinear formula for cFS} for a circular Fermi surface (with two convex critical points) to generically shaped Fermi surfaces by adding up contributions from all critical points selected by the collinear $\bq$'s:
\begin{equation}\label{eq: general Fermi surface_k-space}
    s_3(\{\bq_a\}) =  \frac{\abs{\bq_1\times \bq_2}}{8\pi^2} \sum_p \eta_p\mathcal{G}\left(\abs{\mathbb{C}_p} R_{\{\bq_a\}}\right).
\end{equation}
Here $\mathbb{C}_p$ is the local curvature of the Fermi surface at the critical point $\bk_p$ and $\eta_p = \text{sgn}(\mathbb{C}_p)$ is its signature ($\pm 1$ for an electron-like/hole-like curvature). An application of Eq.~\eqref{eq: general Fermi surface_k-space} is exemplified
 in Fig.~\ref{fig: Fig1}(c) which sums up four convex and two concave critical points. When $\bq_{1,2,3}$ are \textit{finite} and exactly collinear (i.e., $R_{\{\bq_a\}} \rightarrow \infty$), we obtain $s_3 = \abs{\bq_1} \abs{\bq_2} \abs{\bq_3}\sum_p \mathbb{C}_p/(32\pi^2)$ which encodes the geometry of the Fermi surface
 with local information about the critical points selected by the collinear direction. On the other hand, in the LWC limit where $\abs{\mathbb{C}_p}\abs{\bq_a}^2 \ll\abs{q_{\perp}}\ll \abs{\bq_a}$,
 the topological formula in Eq.~\eqref{eq: topological formula, momentum space} is recovered as a sum over critical points as $\sum_p \eta_p = 2\chi_F$ \cite{Kane2022a}. This reinterpretation of the topological formula shows that the $\abs{\bq_1\times\bq_2}$-singularity is sharply defined for a generically shaped Fermi surface under the long-wavelength collinear limit in Eq.~\eqref{eq: q-collinear limit}. 

\noindent {\color{blue}\emph{Long-range collinear correlations--}} Singularities in 
momentum space are connected to long-range correlations in 
real space.  The $\abs{\bq_1\times\bq_2}$-singularity in the LWC limit implies a long-range correlation favoring a collinear configuration with 
$\br_{1,2,3}$ forming a straight line, see Fig.~\ref{fig: Fig1}(d). The straight-line direction in 
real space is perpendicular to the collinear $\bq$'s. From the Fourier transform of Eq.~\eqref{eq: general Fermi surface_k-space} that gives  $\mathfrak{s}_3(\br_{13}, \br_{23})\equiv\cc{\rho(\br_1) \rho(\br_2) \rho(\br_3)}$ (with $\br_{ij}\equiv \br_i- \br_j$), we obtain \cite{supp}
\begin{equation}\label{eq: general Fermi surface_r-space}
\begin{split}
    \mathfrak{s}_3 &= \sum_p \frac{\eta_p}{16\pi^4} \mathcal{R}\left(\abs{\br_{13} \times \br_{23}}, \sqrt{\abs{\mathbb{C}_p}\abs{\br_{13}}\abs{\br_{23}}\abs{\br_{12}}}\right)
\end{split}
\end{equation}
where $\mathcal{R}(x,a) \equiv a^{-3}\sqrt{2/\pi}\sin[x^2/(2a^2)-\pi/4]$ is a regularized form of the second derivative of the delta function $\delta''(x)$ with an effective width $a$ \cite{delta}. Thus the $\abs{\bq_1\times\bq_2}$-singularity 
implies a real space correlation that is dominated by collinear configurations with $\abs{\br_{13}\times \br_{23}} \lesssim \sqrt{\abs{\mathbb{C}_p}\abs{\br_{13}}\abs{\br_{23}}\abs{\br_{12}}}$, with a peak that decays as $\abs{\br_{ij}}^{-3/2}$
for 
a Fermi surface critical point $p$ with velocity $\bv_p \parallel \br_{ij}$. Intuitively in spirit of diffraction, the straight-line correlation is sharper (resp. fuzzier) for a smaller (resp. larger) local Fermi surface curvature $ \abs{\mathbb{C}_p} \equiv k_{F,p}^{-1}$.\\

\begin{figure}
    \centering
    \resizebox{\columnwidth}{!}{\includegraphics[]{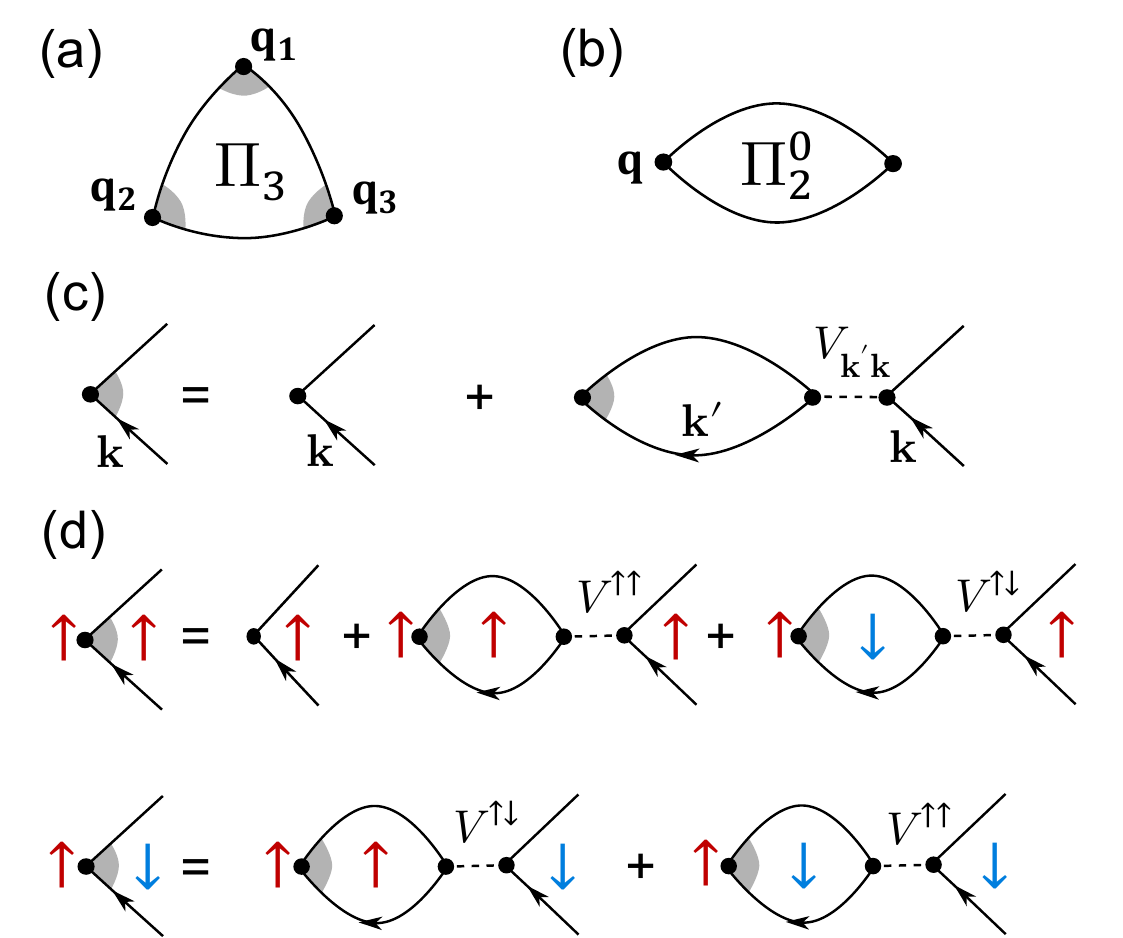}}
    \caption{(a) Three-point bubble $\Pi_3$ dressed with vertex renormalizations. (b) Polarization bubble $\Pi^0_2$. (c) Dyson equation for the vertex renormalization $\Lambda$ in spinless Fermi liquids. (d) Vertex renormalizations, $\Lambda^{\uparrow\uparrow}$ and  $\Lambda^{\uparrow\downarrow}$, in spinful Fermi liquids. Vertices (labeled as dots) represent electron density operators, solid lines represent fermionic propagators, and dashed lines represent renormalized density-density forward-scattering interactions. }
    \label{fig: Fig2}
\end{figure}

\noindent {\color{blue}\emph{Singularity in Fermi liquids--}} We are now in position to study interaction effects on the singularity, making use of the LWC limit in Eq.~\eqref{eq: q-collinear limit}. 
The study of interactions in a Fermi liquid was pioneered by Landau, who formulated a phenomenological model parameterized by the interaction $f({\bf k}',{\bf k})$ between quasiparticles at the Fermi surface \cite{landau1957theory}.  This was justified within a perturbative expansion by noting that in the ${\bf q},\Omega\rightarrow 0$ limit, the exact vertex function $\Gamma(k,k';k+q,k'-q)$ with $k=({\bf k},\omega)$ and $q = ({\bf q},\Omega)$, which includes interactions to all orders in perturbation theory, exhibits a singular dependence on the ratio $\Omega/\abs{\bq}$ \cite{landau1959theory, abrikosovmethods}.   The Landau interaction $f({\bf k}',{\bf k})$ is associated with the $|{\bf q}|/\Omega\rightarrow 0$ limit.   For $|{\bf q}|/\Omega$ finite, $\Gamma$ is obtained by summing a geometric series that resembles a sum of polarization bubble diagrams, and incorporates the singular in $({\bf q},\Omega)$ part of the particle-hole propagator.  

An alternative formulation is to describe this within a low energy effective theory, in which high energy excitations outside of a thin shell $\abs{E-E_F}\lesssim E_\Lambda \ll E_F$ surrounding the Fermi surface are integrated out using the renormalization group \cite{Benfatto1990, Polchinski1992, Shankar1994}. The only interactions that survive to low energy are the marginal forward scattering interactions between particles at ${\bf k}$ and ${\bf k}'$ on the Fermi surface, and those renormalized interactions define the Landau parameters.   In this approach, the geometric sum of bubble diagrams for the vertex renormalization is justified in a manner reminiscent of the large-$N$ expansion \cite{Shankar1994}, where $N \sim E_F/E_\Lambda$ characterizes the number of independent patches on the Fermi surface.   This approach to incorporating the Landau parameters is also built into theories based on bosonization of the Fermi surface \cite{Haldane1994, Houghton1993, houghton2000multidimensional,
Fradkin1994a, Fradkin1994b, Khveshchenko1995, Son2022,mehta2023postmodern}, which amount to performing the bubble summation when the interactions between the bosons are set to zero. 

Here, we adopt the low energy effective fermion theory with a renormalized forward scattering interaction
$\frac{1}{2}\sum_{{\bf k},{\bf k}',{\bf q}}V_{{\bf k}'{\bf k}}c_{{\bf k}'-{\bf q}}^\dagger c_{{\bf k}'}c_{{\bf k}+{\bf q}}^\dagger c_{\bf k}$, where the momenta are restricted to a shell surrounding the Fermi surface.  We compute the three-point correlator $\Pi_3(\{\bq_a, \tau_a\}) = \int\frac{d^2\bq_3}{(2\pi)^2} \cc{T_\tau [\rho_{\bq_1}(\tau_1)\rho_{\bq_2}(\tau_2)\rho_{\bq_3}(\tau_3)]}$ using a zero temperature Euclidean time formalism by evaluating the diagram in Fig.~\ref{fig: Fig2}(a), whose equal-time limit gives  
\begin{equation}
s_3 = \int \frac{d^2\bk}{(2\pi)^2} \left(\prod_{a=1}^3 d\tau_a \Lambda_{\bf k}(\bq_a,\tau_a)\right)
\Pi_{3,\bk}^0(\{\bq_a,\tau_a\}).
\label{s3 integral}
\end{equation}
Here $\Pi_3^0 = \int \frac{d^2\bk}{(2\pi)^2}\Pi^0_{3,\bk}$ is the bare three-point bubble diagram, while interaction effects are captured by the vertex renormalization $\Lambda_{\bf k}(\{\bq,\tau\})$ that incorporates the geometric series of polarization bubbles $\Pi^0_2({\bf q},\tau)$, see Fig.~\ref{fig: Fig2}(b,c).  

An important simplification arises in the LWC limit with $|q_{\perp}| \ll |{\bf q}_a|$, where there is a separation of time scales between $\Pi^0_3$ and $\Pi^0_2$.   The time scale for variation of $\Pi^0_2({\bf q},\tau)$ is set by $(v_F|{\bf q}|)^{-1}$.   We will show below that the time scale for variation of $\Pi^0_3$ is set by $(v_F |q_{\perp}|)^{-1} \gg (v_F|{\bf q}|)^{-1}$.   This is plausible because, as shown in Fig.~\ref{fig: Fig1}(d), the LWC limit corresponds to the real space positions ${\bf r}_{1,2,3}$ that are separated by an amount of order $|q_{\perp}|^{-1}$.  This means that in the LWC limit, 
$\Pi^0_3$ is {\it time independent} as compared with $\Pi_2({\bf q},\tau)$ (and hence $\Lambda_{\bk}(\bq,\tau)$),
so that the time integral of $\Lambda_{\bk}(\bq,\tau)$ in Eq.~\eqref{s3 integral} amounts to evaluating $\Lambda_\bk(\bq,\Omega)$ in the limit $\Omega/v_F|{\bf q}| \rightarrow 0$.
It follows that in the LWC limit, $\Lambda_{\bk} \equiv \int d\tau \Lambda_\bk(\bq\rightarrow 0,\tau)$ satisfies the Dyson equation (Fig.~\ref{fig: Fig2}(c)),
\begin{equation}\label{eq: Dyson}
   \Lambda_{\bk} = 1-\frac{1}{(2\pi)^2}\int_{S_F}d\bk' \Lambda_{\bk'}\frac{V_{\bk'\bk}}{\abs{\bv_{\bk'}}}.
\end{equation}

The analysis of $\Pi_{3,\bk}^0(\{\bq_a,\tau_a\})$ is performed in detail in the supplemental material \cite{supp}.   We find that in the $\bq_a \rightarrow 0$ limit, 
\begin{equation}\label{eq: Pi0_3k}
\begin{split}
    \Pi_{3,\bk}^0&(\{\bq_a, \tau_a\}) = e^{-\bv_\bk\cdot\sum_{a=1}^3\bq_a\tau_a} s_{32}\theta(s_{32}\bv_\bk\cdot\bq_3) \\
    &[\bq_2\cdot\nabla_\bk\theta(\bv_\bk\cdot\bq_{1}) \bq_3\cdot\nabla_\bk f-(2\leftrightarrow 3)]
\end{split}
\end{equation}
where $s_{ab}\equiv \text{sgn}(\tau_{ab})$ and we have kept the leading $q^2$ contribution and dropped regular terms that do not contribute to the singular $\abs{\bq_1\times\bq_2}$-dependence. Notice that $\nabla_\bk f$ and $\nabla_\bk\theta(\bv_\bk \cdot \bq_1)$  restrict $\bk$ to be on the  Fermi surface and at a critical point $\bk_p$ where $\bv_{p} \cdot \bq_1 = 0$.  This result can be simplified by integrating over a neighborhood surrounding each critical point via a change of variables,  $(k_x, k_y) \rightarrow (E_\bk, \bv_\bk\cdot \bq_1)$ with a Jacobian $J_\bk = \hat{z}\cdot[ \bv_\bk\times \nabla_\bk(\bv_\bk\cdot\bq_1)]$, leading to \cite{supp}
\begin{equation}\label{eq: Pi0_3k simplified}
\Pi_{3,\bk}^0 = 
\abs{\bq_1\times\bq_2}
\sum_p \eta_p\delta^2(\bk - \bk_p) \Theta(\bv_p\cdot\sum_p \bq_a\tau_a)
\end{equation}
with $\eta_p$ the signature of  
$\bk_p$ and   $\Theta(x)\equiv\theta(x)e^{-x}$.

The time-dependent factor then becomes $e^{- v_p\abs{q_{\perp}\tau_{23}}}$, confirming our previous assertion about the time scale of $\Pi^0_3$ and justifies our treatment of the time-integral in Eq.~\eqref{s3 integral}.
Moreover, since ${\rm sgn} (\bv_p\cdot\sum_a \bq_a\tau_a) ={\rm sgn}(\bv_p\cdot\bq_2 \tau_{23})$ is odd under the interchange of $\tau_2$ and $\tau_3$, while the rest of the integrand in Eq.~\eqref{s3 integral} is even, we may replace $\theta(\bv_p\cdot\sum_a \bq_a\tau_a)$ by $1/2$.   We thus obtain for generically shaped Fermi surfaces and in the presence of Fermi-liquid interactions,
\begin{equation}\label{eq: general FS interaction effect for s3}
    s_3(\{\bq_a\}) = \frac{\abs{\bq_1\times \bq_2}}{8\pi^2}\sum_p \eta_p\Lambda^3_{\bk_p},
\end{equation}
up to regular terms \cite{regular}. For $\Lambda_{\bf k}=1$ we recover the critical point formulation of Eq.~\eqref{eq: topological formula, momentum space}.   For $\Lambda_{\bf k} \ne 1$, we have established that the $\abs{\bq_1\times \bq_2}$-singularity in $s_3$ survives the presence of interactions, as a universal feature of Fermi liquids. The effect of interactions is to renormalize the coefficient of singularity.

While it has been pointed out that higher-order interactions can generate order-$\abs{\bq}^2$ correction to $s_3$ \cite{Son2022}, we have explicitly checked (see the supplementary material \cite{supp}) for a three-body density-density interaction that no such singular $\abs{\bq_1\times \bq_2}$ term is generated. Generally, we expect that irrelevant interactions (in the renormalization group sense) do not contribute to the $\abs{\bq_1\times \bq_2}$-singularity in $s_3$ which is tied to the long-range collinear correlation highlighted above.

We now circle back to a circular electron-like Fermi surface in the presence of an isotropic interaction, which can be expanded as $V_{\bk'\bk} = \sum_{m=0}^\infty V_m\cos{m \theta_{\bk'\bk}} $, with $\theta_{\bk'\bk}$ the angle between the forward-scattered Fermi points. Equation \eqref{eq: Dyson} reduces to $\Lambda = [1+V_0 k_F/(2\pi v_F)]^{-1} = (1+F_0)^{-1}$, where we have introduced the Landau parameter $F_0 = \mathcal{N}_F V_0$ (with $\mathcal{N}_F = k_F/(2\pi v_F)$ the density of states at the Fermi surface). Hence, up to regular terms,
\begin{equation}\label{eq: main result spinless}
    s_3(\{\bq_a\}) = \frac{\abs{\bq_1\times\bq_2}}{(2\pi)^2} \frac{1}{(1+F_0)^3}.
\end{equation}

\noindent {\color{blue}\emph{Spinful Fermi liquids--}} Next we extend the previous results for spinless fermions to two-component fermions (labeled $\uparrow$ and $\downarrow$). 
We assume the two components behave identically, and consider general effective interactions 
of the form
$\frac{1}{2}\sum_{{\bf k},{\bf k}',{\bf q},\sigma=\uparrow,\downarrow}\big[V^{\uparrow\uparrow}_{{\bf k}'{\bf k}}c_{\sigma, {\bf k}'-{\bf q}}^\dagger c_{\sigma,{\bf k}'}c_{\sigma, {\bf k}+{\bf q}}^\dagger c_{\sigma, \bk} +V^{\uparrow\downarrow}_{\bk'\bk} c_{\sigma, {\bf k}'-{\bf q}}^\dagger c_{\sigma,{\bf k}'}c_{\bar{\sigma}, {\bf k}+{\bf q}}^\dagger c_{\bar{\sigma}, \bk}\big]$,
where $V^{\uparrow\uparrow}$ and $V^{\uparrow\downarrow}$ represent, respectively, the intra-component and inter-component renormalized forward scattering interactions. Correspondingly, there are two types of vertex renormalizations, $\Lambda^{\uparrow\uparrow}$ and $\Lambda^{\uparrow\downarrow}$, as shown in Fig.~\ref{fig: Fig2}(d), which satisfy the following Dyson equations: $\Lambda^{\uparrow\uparrow}_\bk = 1-\int_{S_F} \frac{d\bk'}{(2\pi)^2}\left(\Lambda^{\uparrow\uparrow}_{\bk'}V^{\uparrow\uparrow}_{\bk'\bk}+\Lambda^{\uparrow\downarrow}_{\bk'}V^{\uparrow\downarrow}_{\bk'\bk}\right)/\abs{\bv_{\bk'}}$ and $\Lambda^{\uparrow\downarrow}_{\bk} = -\int_{S_F}\frac{d\bk'}{(2\pi)^2} \left(\Lambda^{\uparrow\uparrow}_{\bk'}V^{\uparrow\downarrow}_{\bk'\bk}+\Lambda^{\uparrow\downarrow}_{\bk'}V^{\uparrow\uparrow}_{\bk'\bk}\right)/\abs{\bv_{\bk'}}$. We now repeat the analysis around Eq.~\eqref{eq: general FS interaction effect for s3} for two-component Fermi liquids. We focus on two quantities of interest:  (i)
the single-component (or
same-spin) density correlation $s_3^{\uparrow}(\bq_1,\bq_2) \equiv \int\frac{d^2\bq}{(2\pi)^2}\cc{\rho^\uparrow_{\bq_1}\rho^\uparrow_{\bq_2}\rho^\uparrow_\bq}$, and (ii) the total density correlation $s_3$ with $\rho = \rho^\uparrow +\rho^\downarrow$. 
For $s_3^{\uparrow}$, we simply replace $\Lambda^3_{\bk} \rightarrow (\Lambda^{\uparrow\uparrow}_{\bk})^3+(\Lambda^{\uparrow\downarrow}_{\bk})^3$ as the bubble in Fig.~\ref{fig: Fig2}(a) can either have the same or opposite spin as the density operators. As for $s_3$, we also need to consider $\cc{\rho^{\uparrow}\rho^\uparrow\rho^\downarrow}$ and $\cc{\rho^{\downarrow}\rho^\downarrow\rho^\uparrow}$ (together with permutations), and for each of these terms we 
replace $\Lambda^3_{\bk} \rightarrow (\Lambda^{\uparrow\uparrow}_{\bk})^2\Lambda^{\uparrow\downarrow}_\bk+(\Lambda^{\uparrow\downarrow}_{\bk})^2\Lambda^{\uparrow\uparrow}_\bk$.

For a circular Fermi surface ($\chi_F=1$ per spin) with an isotropic interaction $V^{\uparrow\uparrow/\uparrow\downarrow}_{\bk'\bk}=\sum_{m=0}^\infty V^{\uparrow\uparrow/\uparrow\downarrow}_m \cos m\theta_{\bk'\bk}$, again only the $0$-th component matters to the vertex renormalizations. Introducing the spin-symmetric and spin-asymmetric Landau parameters,
$F^{s/a}_m \equiv \mathcal{N}_F(V^{\uparrow\uparrow}_m \pm V^{\uparrow\downarrow}_m)$, we obtain $\Lambda^{\uparrow\uparrow} = \left[1+\frac{1}{2}(F^s_0+F^a_0)\right](1+F^s_0)^{-1}(1+F^a_0)^{-1}$ and $\Lambda^{\uparrow\downarrow} = \frac{1}{2}(F^a_0-F^s_0)(1+F^s_0)^{-1}(1+F^a_0)^{-1}$. Thus, up to regular terms, we arrive at
\begin{subequations}\label{eq: main result spinful}
    \begin{align}
        s^{\uparrow}_3 &= \frac{\abs{\bq_1\times\bq_2}}{(2\pi)^2} \frac{(1+F^a_0)^2+3(1+F^s_0)^2}{4(1+F^s_0)^3(1+F^a_0)^2 }, \\
        s_3 &= \frac{\abs{\bq_1\times\bq_2}}{(2\pi)^2}\frac{2}{(1+F^s_0)^3}.
    \end{align}
\end{subequations}
Note that Eq.~\eqref{eq: main result spinful} holds even if $F_0^{s,a}$ are not small.

\noindent {\color{blue}\emph{Experimental implications--}} Finally, we discuss implications for quantum gas experiments \cite{daix2025probing}, where the microscopic interaction is a contact interaction between opposite spins. 
We use results \cite{tarik}
derived in Ref. \cite{Randeria1992} which perturbatively relate the Landau parameters to second order in an interaction parameter: $F^s_0 = \mathcal{I}+ \mathcal{I}^2(2-\ln2)$ and $F^a_0 = -\mathcal{I}+\mathcal{I}^2\ln 2$, where $\mathcal{I} \equiv -1/\ln(k_Fa)$ and $a$ is the scattering length. Substituting these into Eq.~\eqref{eq: main result spinful} and expanding to order $\mathcal{I}^2$:
\begin{subequations}
    \begin{align}
        s^{\uparrow}_3 &= \frac{\abs{\bq_1\times\bq_2}}{(2\pi)^2} \left[1+\mathcal{O}(\mathcal{I}^3)\right], \\
      s_3 &=  2\frac{\abs{\bq_1\times\bq_2}}{(2\pi)^2} \left[1-3\mathcal{I}+3\mathcal{I}^2\ln 2+\mathcal{O}(\mathcal{I}^3)\right].  
    \end{align}
\end{subequations}
The vanishing correction to $s_3^\uparrow$ at order $\mathcal{I}$ is expected for the contact interaction between opposite spins. The vanishing correction at order $\mathcal{I}^2$, however, is a less
trivial prediction. This robustness of $s^\uparrow_3$ was first observed in a recent experiment reported in Ref.~\cite{daix2025probing}, where the free Fermi gas prediction 
matched  the experimental result for moderately strong attractive interactions up to
$\mathcal{I}\sim -0.5$. Our theory provides a preliminary explanation for this phenomenon and motivates an in-depth comparison between the spin-resolved correlation and the total density correlation in future experiments. 

\noindent {\color{blue}\emph{Conclusion--}}
We have studied the equal-time three-point density correlation and established that it generically contains an $\abs{\bq_1\times \bq_2}$-singularity in two dimensional interacting Fermi liquids. This singularity is sharply defined for arbitrary Fermi surface shapes in the long-wavelength collinear limit. The coefficient of this singularity probes the effective interactions at critical points on the Fermi surface. For isotropic Fermi liquids, the singularity is related simply to Landau parameters, and should be testable in quantum gas experiments. An important future direction is to investigate the fate of this singularity in both charged Fermi liquids with long-range interactions and strongly correlated non-Fermi liquids.  

\begin{acknowledgments}
\emph{Acknowledgments.} We are grateful to Tarik Yefsah, Cyprien Daix, and Bruno Peaudecerf for inspiring discussions and collaboration on related projects. We also appreciate Luca Delacrétaz and Mohit Randeria for helpful discussions. 
\end{acknowledgments}

\bibliographystyle{apsrev4-1.bst}
\bibliography{references}

\clearpage
\newpage
\widetext
\begin{center}
\textbf{\large Supplemental Materials for \\ ``Singular three-point density correlations in two-dimensional Fermi liquids"}\\
\vspace{0.5cm}
\text{Pok Man Tam and Charles Kane}
\end{center}
\maketitle
\onecolumngrid
\setcounter{secnumdepth}{2}
\renewcommand{\theequation}{\thesection.\arabic{equation}}
\renewcommand{\theHequation}{\theHsection.\arabic{equation}}
\renewcommand{\thefigure}{\thesection.\arabic{figure}} 

The supplemental information consists of three sections. In Sec.~\ref{sec: real space correlation} we provide details on the long-range correlation that favors straight-line configurations. In Sec.~\ref{supp_sec: 3-leg bubble}, we provide details for deriving the three-point density correlation in Fermi liquids. In Sec.~\ref{supp_sec: 3-bdy interaction}, we demonstrate that the irrelevant three-body density interaction does not generate any $\abs{\bq_1\times\bq_2}$-singularity in the equal-time three-point density correlation. 

\section{Real-space correlation}\label{sec: real space correlation}
\setcounter{equation}{0}
\setcounter{figure}{0} 
Here we provide the derivation for Eqs. \eqref{eq: general Fermi surface_r-space} in the main text, which reflect the real-space long-range straight-line correlation in a Fermi gas. Let us begin with a circular electron-like Fermi surface (with Fermi radius $k_F$), the three-point density correlator in the momentum space takes the following form (c.f. Eq. \eqref{eq: collinear formula for cFS}):
\begin{equation}
    s_3(\bq_1, \bq_2) = \frac{q_1}{(2\pi)^2} \begin{cases}
         \abs{q_{\perp}}  &\text{for}\; \abs{q_{\perp}} \geq q_0\\
        \frac{1}{2}(q_0+q^2_{2\perp}/q_0) &\text{for}\; \abs{q_{\perp}} \leq q_0
    \end{cases}
\end{equation}
where we have expressed
\begin{equation}
    \bq_1 = q_1 \hat{\bq}_1, \quad\bq_2 = q_{\perp}(\hat{\bq}_1\times \hat{z})+q_{2\parallel}\hat{\bq}_1,\quad\text{and}\quad q_0 = \frac{1}{2k_F}\abs{q_{2\parallel}(q_1+q_{2\parallel})}.
\end{equation}
We have chosen to focus on the collinear limit in the momentum space, which as explained in the main text is related to the straight-line correlation in the real space. The real-space correlation $\mathfrak{s}_3(\br_1, \br_2) \equiv \cc{\rho(\br_1) \rho(\br_2) \rho(0)}$ is obtained via the Fourier transform:
\begin{equation}
    \mathfrak{s}_3(\br_1, \br_2) = \frac{1}{(2\pi)^4} \int d^2\bq_1 d^2\bq_2\;s_3(\bq_1, \bq_2) e^{i(\bq_1\cdot\br_1+\bq_2\cdot \br_2)} = \frac{1}{(2\pi)^6} R(\br_1, \br_2)
\end{equation}
with 
\begin{align}
    R(\br_1, \br_2) &=  \int d^2\bq_1 \int dq_{2 \parallel}\;2q_1 e^{i\bq_1\cdot\br_1+i q_{2\parallel} r_{2\parallel}} \Big[\int_{0}^{q_0}\frac{dq_{\perp}}{2}(q_0+\frac{q^2_{\perp}}{q_0}) \cos{(q_{\perp} r_{2\perp})}+\int_{q_0}^\infty dq_{\perp}\;q_{\perp} \cos(q_{\perp} r_{2\perp})\Big]\\
    &= - \int d^2\bq_1 e^{i\bq_1\cdot\br_1} \frac{2q_1}{r^2_{2\perp}}\int dq_{2 \parallel}\;e^{iq_{2\parallel} r_{2\parallel}}\frac{\sin{(q_0 r_{2\perp})}}{q_0 r_{2\perp}}, 
\end{align}
where $r_{2\parallel}\equiv \br_2\cdot\hat{\bq}_1$ and $r_{2\perp}\equiv - \br_2\cdot(\hat{z}\times\hat{\bq}_1)$.
To obtain the last equality, we have used $\frac{1}{2}\int_0^X du\; (X+u^2/X)\cos(u) = \cos(X)+(X-1/X)\sin(X)$, and $\lim_{\epsilon\rightarrow0^+}\int_X^\infty du\;u \cos(u) e^{-\epsilon u}  = -\cos(X)-X\sin(X)$. Next, we rewrite $\frac{\sin(X)}{X} = \frac{1}{2}\int_{-1}^1 d\mu \;e^{i\mu X}$ so that the $q_{2\parallel}$-integral can be performed as a Fresnel integral (a complex Gaussian integral):
\begin{align}
    R(\br_1, \br_2) 
    &=- \int_{-1}^1 d\mu\int d^2\bq_1 e^{i\bq_1\cdot\br_1} \frac{q_1}{r^2_{2\perp}}\int dq_{2 \parallel}\;e^{i \frac{\mu r_{2\perp}}{2k_F}(q^2_{2\parallel}+q_1q_{2\parallel})+iq_{2\parallel} r_{2\parallel}}\\
    &= - \int_{-1}^1 d\mu\int d^2\bq_1 e^{i\bq_1\cdot\br_1} \frac{q_1}{r^2_{2\perp}}\sqrt{\frac{2\pi k_F}{\abs{\mu r_{2\perp}}}}e^{\text{sgn}(\mu r_{2\perp})i\frac{\pi}{4}-i\frac{k_F}{2\mu r_{2\perp}}(r_{2\parallel}+\frac{\mu r_{2\perp} q_1}{2k_F})^2} \\
    & = - \int_{-1}^1 d\mu\int d^2\bq_1 e^{i\bq_1\cdot\br_1} q_1 \sqrt{\frac{2\pi k_F}{\abs{\mu r^5_{2\perp}}}}e^{\text{sgn}(\mu r_{2\perp})i\frac{\pi}{4}-i\frac{k_F r_{2\parallel}^2}{2\mu r_{2\perp}}-i\frac{q_1 r_{2\parallel}}{2}-i\frac{\mu q_1^2 r_{2\perp}}{8k_F}}
\end{align}
Next, we reparametrize as follows:
\begin{equation}
    \bq_1 = q_{1\parallel} \hat{\br}_2 - q_{1\perp} (\hat{z}\times \hat{\br}_2) \implies \begin{cases}
r_{2\parallel} & = r_2 q_{1\parallel}/q_1 \\
r_{2\perp} & = -r_2 q_{1\perp}/q_1
    \end{cases}, \quad \bq_1\cdot \br_1 = q_{1\parallel}r_{1\parallel}+q_{1\perp}r_{1\perp} \;\text{with}\;\begin{cases}
        r_{1\parallel} &= \br_1 \cdot \hat{\br}_2\\
        r_{1\perp}  &= -\br_1\cdot(\hat{z}\times \hat{\br}_2)
    \end{cases}
\end{equation}
where $r_2 \equiv\abs{\br_2}$ (and similarly for other quantities). Then
\begin{equation}
    R(\br_1, \br_2)  = - \int_{-1}^1 d\mu\int d^2\bq_1 e^{iq_{1\parallel}r_{1\parallel}+iq_{1\perp}r_{1\perp}} \sqrt{\frac{2\pi k_F q_1^7}{r_2^5\abs{\mu q_{1\perp}^5}}}e^{-\text{sgn}(\mu q_{1\perp})i\frac{\pi}{4}+i\frac{k_F r_2 q_{1\parallel}^2 }{2\mu q_1 q_{1\perp} }-i\frac{r_2 q_{1\parallel}}{2}+i\frac{\mu r_2 q_1 q_{1\perp}}{8k_F}}
\end{equation}
As we are interested in the long-range correlation, i.e. $k_F r_{1,2} \gg 1$, the phase factor $\exp(i\frac{k_F r_2 q_{1\parallel}^2 }{2\mu q_1 q_{1\perp} })$ in the integrand above suggests that the integral is dominated by $q^2_{1\parallel}/(q_1 \abs{q_{1\perp}}) \ll 1 \implies \abs{q_{1\parallel}} \ll \abs{q_{1\perp}} \implies q_1 \approx \abs{q_{1\perp}}$. With this replacement, the $q_{1\parallel}$-integral can first be easily performed, followed by the $\mu$-integral and lastly by the $q_{1\perp}$-integral:
\begin{align}
    &R(\br_1, \br_2)  = - \int_{-1}^1 d\mu\int d q_{1\perp} dq_{1\parallel} e^{iq_{1\parallel}r_{1\parallel}+iq_{1\perp}r_{1\perp}} \sqrt{\frac{2\pi k_F q_{1\perp}^2}{r_2^5\abs{\mu}}}e^{-\text{sgn}(\mu q_{1\perp})i\frac{\pi}{4}+i\frac{k_F r_2 q_{1\parallel}^2 }{2\mu \abs{q_{1\perp}} q_{1\perp} }-i\frac{r_2 q_{1\parallel}}{2}+i\frac{\mu r_2 \abs{q_{1\perp}} q_{1\perp}}{8k_F}} \\
    & = - \int_{-1}^1 d\mu\int d q_{1\perp}\sqrt{\frac{2\pi k_F q_{1\perp}^2}{r_2^5\abs{\mu}}}e^{iq_{1\perp}r_{1\perp}-\text{sgn}(\mu q_{1\perp})i\frac{\pi}{4}+i\frac{\mu r_2 \abs{q_{1\perp}} q_{1\perp}}{8k_F}} \int dq_{1\parallel}e^{i\frac{k_F r_2 q_{1\parallel}^2 }{2\mu \abs{q_{1\perp}} q_{1\perp} }+i(r_{1\parallel}-\frac{r_2}{2})q_{1\parallel}} \\
    &= - \int_{-1}^1 d\mu\int d q_{1\perp}\sqrt{\frac{2\pi k_F q_{1\perp}^2}{r_2^5\abs{\mu}}}e^{iq_{1\perp}r_{1\perp}-\text{sgn}(\mu q_{1\perp})i\frac{\pi}{4}+i\frac{\mu r_2 \abs{q_{1\perp}} q_{1\perp}}{8k_F}} e^{\text{sgn}(\mu q_{1\perp})i\frac{\pi}{4}-i\frac{\mu\abs{q_{1\perp}}q_{1\perp}}{2k_F r_2}(r_{1\parallel}-\frac{r_2}{2})^2}\sqrt{\frac{2\pi\abs{\mu}q_{1\perp}^2}{k_F r_2}} \\
    & = - \int_{-1}^1 d\mu\int d q_{1\perp}\frac{2\pi q_{1\perp}^2}{r_2^3} e^{i q_{1\perp} r_{1\perp} -i\frac{\mu\abs{q_{1\perp}}q_{1\perp}}{2k_F r_2}r_{1\parallel}(r_{1\parallel}-r_2) } \\
    & = - \int d q_{1\perp}\frac{4\pi q_{1\perp}^2}{r_2^3} e^{i q_{1\perp} r_{1\perp}} \frac{\sin(\frac{q^2_{1\perp}\abs{r_{1\parallel}(r_{1\parallel}-r_2)}}{2k_F r_2})}{\frac{q^2_{1\perp}\abs{r_{1\parallel}(r_{1\parallel}-r_2)}}{2k_F r_2}}\\
    & = - \frac{8\pi k_F}{r^2_2\abs{r_{1\parallel}(r_{1\parallel}-r_2)}}\Im \Big[\int dq_{1\perp} e^{i\frac{\abs{r_{1\parallel}(r_{1\parallel}-r_2)}}{2k_F r_2}q^2_{1\perp}+ir_{1\perp}q_{1\perp}}\Big]\\
    & = - \frac{8\pi k_F}{r^2_2\abs{r_{1\parallel}(r_{1\parallel}-r_2)}}\Im \Big[ e^{i\frac{\pi}{4}-i\frac{k_F r_2 r_{1\perp}^2}{2\abs{r_{1\parallel}(r_{1\parallel}-r_2)}}} \sqrt{\frac{2\pi k_F r_2}{\abs{r_{1\parallel}(r_{1\parallel}-r_2)}}}\Big]\\
    & = 4 \Big[\frac{2\pi k_F}{r_2 \abs{r_{1\parallel}(r_{1\parallel}-r_2)}}\Big]^{\frac{3}{2}} \sin \Big[\frac{k_F r_2}{2\abs{1-\frac{r_2}{r_{1\parallel}}}} \frac{r_{1\perp}^2}{r_{1\parallel}^2} - \frac{\pi}{4}\Big].
\end{align}
The last expression bears a very simple interpretation in the long-range limit: when both $r_1$ and $r_2$ tend to infinity, $\frac{k_F r_2}{2\abs{1-\frac{r_2}{r_{1\parallel}}}} \rightarrow \infty$, and hence to avoid cancellation from the infinitely oscillatory sine function, it effectively imposes the constraint $\abs{r_{1\perp}} \ll \abs{r_{1\parallel}}$. Thus, 
\begin{align}
    \mathfrak{s}_3(\br_1, \br_2) &= \frac{1}{(2\pi)^6}\cdot 4 \Big[\frac{2\pi k_F}{\abs{\br_1}\abs{\br_2}\abs{\br_1-\br_2}}\Big]^{\frac{3}{2}} \sin \Big[\frac{k_F \abs{\br_1\times\br_2}^2}{2\abs{\br_1}\abs{\br_2}\abs{\br_1-\br_2}} - \frac{\pi}{4}\Big]
    = \frac{1}{8\pi^4}\mathcal{R}\left(\abs{\br_1 \times \br_2}, \sqrt{k_F^{-1}\abs{\br_1}\abs{\br_2}\abs{\br_{12}}}\right)
\end{align}
with $\br_{12} \equiv \br_1 - \br_2$ and 
\begin{equation}
    \mathcal{R}(x,a) \equiv \frac{1}{a^3}\sqrt{\frac{2}{\pi}}\sin(\frac{x^2}{2a^2}-\frac{\pi}{4}).
\end{equation}
We have derived the above result assuming a circular electron-like Fermi surface, i.e.,
\begin{equation}
    s_3(\{\bq_a\}) = \frac{1}{(2\pi)^2}\abs{\bq_1\times \bq_2}\cdot\mathcal{G}\left(k_F^{-1}R_{\{\bq_a\}}\right)\;,\;\; \mathcal{G}(X)=\begin{cases}
    1 &\text{if }X\leq1,\\
    \frac{1}{2X}(1+X^2) &\text{if }X\geq 1.
\end{cases}
\end{equation}
For a generically shaped Fermi surface, as we have argued in the main text, the momentum-space density correlation $s_3$ is dominated by Fermi surface critical points (characterized by local curvature $\mathbb{C}_p$ and signature $\eta_p$) as
\begin{equation}
    s_3(\{\bq_a\}) =  \frac{\abs{\bq_1\times \bq_2}}{8\pi^2} \sum_p \eta_p\mathcal{G}\left(\abs{\mathbb{C}_p} R_{\{\bq_a\}}\right),
\end{equation}
and hence the Fourier transform into the real-space correlation is straightforward:
\begin{equation}
    \mathfrak{s}_3(\br_1, \br_2) = \frac{1}{16\pi^4}\sum_p \eta_p \mathcal{R}\left(\abs{\br_1 \times \br_2}, \sqrt{\abs{\mathbb{C}_p}\abs{\br_1}\abs{\br_2}\abs{\br_{12}}}\right).
\end{equation}
Replacing $\br_{1}\rightarrow \br_1-\br_3$ and $\br_{2}\rightarrow \br_2-\br_3$, we arrive at Eq. \eqref{eq: general Fermi surface_r-space} in the main text.\\

Finally, we can use the diffraction picture to give a simple derivation of Eq. \eqref{eq: general Fermi surface_r-space} in the case of a circular Fermi surface. There, the single-particle Green's function is $g(\br)\equiv\langle\psi^\dagger(\br)\psi(\br)\rangle = \int\frac{d^2\bk}{2\pi}e^{i\bk\cdot\br}\theta(E_F-E_\bk) = \frac{k_F}{2\pi \abs{\br}}J_1(k_F \abs{\br})$, where $J_1(x)$ is the Bessel function of the first kind with the asymptotic form $J(x \gg 1) \rightarrow \sqrt{\frac{2}{\pi x}}\sin(x-\frac{\pi}{4})$. In the long-range limit with $k_F\abs{\br_{ij}}\gg 1$, $\mathfrak{s}_3(\{\br_{ij}\}) = g(\br_{12})g(\br_{23})g(\br_{31})+ g(\br_{13})g(\br_{32})g(\br_{21})$ becomes
\begin{equation}
    \mathfrak{s}_3 = \frac{1}{2\pi^4}\sqrt{\frac{2}{\pi}}\left(\frac{k_F}{\abs{\br_{12}}\abs{\br_{23}}\abs{\br_{31}}}\right)^{\frac{3}{2}}\prod_{i\neq j}\sin(k_F\abs{\br_{ij}}-\frac{\pi}{4}).
\end{equation}
Converting the above product into a sum of sines, the long-range correlation is contributed by terms of the form $\sin[k_F(\abs{\br_{12}}+\abs{\br_{23}}-\abs{\br_{31}})-\frac{\pi}{4}]$, together with permutations of $\{12, 23, 31\}$. At most one of these terms can ever be non-vanishing, and this only happens when $\{\br_1, \br_2, \br_3\}$ are almost collinear (so as to avoid a wildly oscillating sine function). In the case depicted in Fig. \ref{fig: Fig1}(d), $\sin[k_F(\abs{\br_{12}}+\abs{\br_{31}}-\abs{\br_{23}})-\frac{\pi}{4}]$ dominates, and Eq. \eqref{eq: general Fermi surface_r-space} is reproduced upon expanding $\abs{\br_{12}}+\abs{\br_{31}}-\abs{\br_{23}}$ to the leading order in the shortest height of the real-space triangle formed by $\{\br_{ij}\}$.

\section{Three-point bubble}\label{supp_sec: 3-leg bubble}
\setcounter{equation}{0}
\setcounter{figure}{0} 

\begin{figure}[H]
    \centering
    \resizebox{\columnwidth}{!}{\includegraphics[]{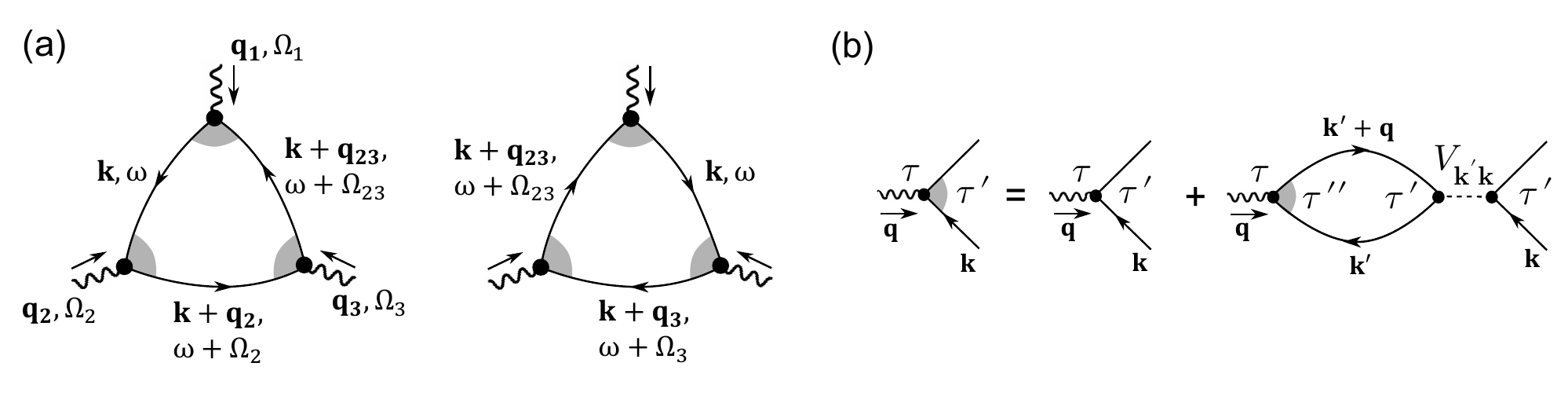}}
    \caption{Feynman diagrams for calculating the three-point density correlation in a Fermi liquid. (a) The three-point bubble $\Pi_3$ is dressed with vertex renormalizations (indicated as shaded vertices) that encode interaction effects. $\Pi_3$ (as well as its non-interacting version $\Pi^0_3$) is contributed by two diagrams and correspond respectively to the two terms in Eq.~\eqref{supp_eq: Pi0_3 from diagrams} and Eq.~\eqref{supp_eq: Pi_3 from diagrams}.  (b) The vertex renormalization $\Lambda_\bk(\bq,\tau'-\tau)$ obeys a Dyson equation, see Eq.~\eqref{supp_eq: Dyson equation in time domain}, where interaction effects are accounted for by summing a geometric series of polarization bubble diagrams. $V_{\bk'\bk}$ is the effective forward scattering interaction and represented as a dashed line.}
    \label{supp_fig: Fig1}
\end{figure}

In this section we provide more details on the calculation of the three-point density correlation in interacting Fermi liquids, and particularly, derive Eqs.~\eqref{eq: Pi0_3k} and ~\eqref{eq: Pi0_3k simplified} in the main text. We focus on spinless Fermi liquids, as generalizations to the spinful case have been discussed in the main text. We use the Matsubara formalism (taking the zero-temperature limit $\beta \rightarrow \infty$) and work with imaginary time. The quantity of interest is
\begin{equation}\label{supp_eq: Pi_3}
    \Pi_3(\{\bq_a, \tau_a\}) = \int \frac{d^2\bq_3}{(2\pi)^2} \cc{T_\tau [\rho_{\bq_1}(\tau_1)\rho_{\bq_2}(\tau_2)\rho_{\bq_3}(\tau_3)]},
\end{equation}
where $\rho_\bq(\tau) = e^{H\tau} \rho_\bq e^{-H\tau}$, with $H$ the Hamiltonian for an interacting Fermi liquid, and $T_\tau$ represents the time-ordered product with $\tau$ increasing from right to left. Notice that $\Pi_3$ is a function of time-differences (depending on $\tau_2-\tau_1$ and $\tau_3-\tau_1$), while momentum conservation dictates that only $\bq_3=-\bq_1-\bq_2$ would contribute. For this analysis of interacting Fermi liquids, we will focus on the long-wavelength collinear (LWC) limit (see Eq.~\eqref{eq: q-collinear limit} in the main text), and focus on extracting the 2nd-order non-analytic $\bq$-dependence of the form $\abs{\bq_1\times \bq_2}$. In the absence of interaction, the three-point density correlation in Eq.~\eqref{supp_eq: Pi_3} corresponds to a bare three-point bubble diagram which we denote as $\Pi_3^0$:
\begin{equation}\label{supp_eq: Pi0_3 from diagrams}
\begin{split}
    \Pi^0_3(\{\bq_a, \tau_a\}) &= \int\frac{d^2\bk}{(2\pi)^2} \Pi^0_{3,\bk}(\{\bq_a, \tau_a\}),\\
    \Pi^0_{3,\bk}(\{\bq_a, \tau_a\}) &=
    \frac{1}{\beta^3}\sum_{\omega, \Omega_{2,3}} \frac{ e^{-i\Omega_2\tau_{21}-i\Omega_{3}\tau_{31}}}{(i\omega-\xi_\bk)(i\omega+i\Omega_2-\xi_{\bk+\bq_2})(i\omega+i\Omega_{23}-\xi_{\bk+\bq_{23}})} + \left(2\leftrightarrow 3\right),
\end{split}
\end{equation}
with $\xi_\bk \equiv E_\bk-E_F$. The first (second) term in $\Pi^0_{3,\bk}$ corresponds to the left (right) diagram in Fig.~\ref{supp_fig: Fig1}(a) \footnote{Notice that the $(-1)$ factor from the fermionic loop cancels with another factor of $(-1)^3$ from the minus sign in the definition of each Green's function.}. Here, $\omega$ corresponds to a fermionic Matsubara frequency, while $\Omega_{2,3}$ are bosonic Matsubara frequencies. We have also introduced the short-hand notations: $\tau_{ab} \equiv \tau_a-\tau_b$, $\Omega_{ab}\equiv \Omega_a+\Omega_b$ and $\bq_{ab}\equiv \bq_a+\bq_b$. We will compute $\Pi^0_{3,\bk}$ shortly, but before that let us first make some remarks about the effect of Fermi-liquid interactions.\\

We adopt the low energy effective fermion theory with a renormalized forward scattering interaction 
\begin{equation}
    \frac{1}{2}\int'\frac{d^2\bk d^2\bk'd^2\bq}{(2\pi)^6}\;V_{{\bf k}'{\bf k}}\;c_{{\bf k}'-{\bf q}}^\dagger c_{{\bf k}'}c_{{\bf k}+{\bf q}}^\dagger c_{\bf k},
\end{equation}
where $\int'$ indicates a constrained integral such that all the momenta are restricted to a shell surrounding the Fermi surface. Here $V_{\bk'\bk}$
parametrizes the effective forward scattering interaction between two patches on the Fermi surface labeled by $\bk'$ and $\bk$, with 
a small momentum transfer $\bq$ between the patches.
The effect of Fermi-liquid interaction on $\Pi_3$ is accounted for by dressing each density vertex with a geometric series of polarization bubbles to obtain an effective vertex renormalization denoted as $\Lambda_\bk$. As such, $\Pi_3(\{\bq_a, \tau_a\}) = \int\frac{d^2\bk}{(2\pi)^2} \Pi_{3,\bk}(\{\bq_a, \tau_a\})$ and
\begin{equation}\label{supp_eq: Pi_3 from diagrams}
\begin{split}
    \Pi_{3,\bk} &=
    \frac{1}{\beta^3}\sum_{\omega, \Omega_{2,3}} \int \prod_{a=1}^3 d\tau_a' \frac{\Lambda_{\bk-\bq_1}(\bq_1, \tau'_1-\tau_1) \Lambda_{\bk}(\bq_2, \tau'_2-\tau_2) \Lambda_{\bk+\bq_2}(\bq_3, \tau'_3-\tau_3) e^{-i\Omega_2\tau'_{21}-i\Omega_{3}\tau'_{31}}}{(i\omega-\xi_\bk)(i\omega+i\Omega_2-\xi_{\bk+\bq_2})(i\omega+i\Omega_{23}-\xi_{\bk+\bq_{23}})} + \left(2\leftrightarrow 3\right).
\end{split}
\end{equation}
The above expression can be simplified in the long-wavelength limit, where we can replace $\Lambda_{\bk+\bq}$ by $\Lambda_\bk$
(since to leading order in $\bq$, $V_{\bk',\bk+\bq}=V_{\bk',\bk}$). Thus
\begin{equation}\label{supp_eq: Pi3 with vertex correction, time domain}
    \Pi_3(\{\bq_a,\tau_a\})=\int\frac{d^2\bk}{(2\pi)^2}\left(\int\prod_{a=1}^3 d\tau_a' \;\Lambda_\bk(\bq_a, \tau'_a-\tau_a)\right) \Pi^0_{3,\bk}(\{\bq_a, \tau'_a\}).
\end{equation}
Taking the equal-time limit, with $\tau_a=0$, we obtain Eq.~\eqref{s3 integral} in the main text. \\

The vertex renormalization satisfies the following Dyson equation (Fig.~\ref{supp_fig: Fig1}(b)),
\begin{equation}\label{supp_eq: Dyson equation in time domain}
    \Lambda_\bk (\bq, \tau'-\tau) = \delta(\tau'-\tau)-\int\frac{d^2\bk'}{(2\pi)^2}\int d\tau'' \Lambda_{\bk'}(\bq, \tau''-\tau)\Pi^0_{2, \bk'}(\bq,\tau''-\tau')V_{\bk'\bk},
\end{equation}
where the polarization bubble dressing the $\bq$-vertex is
\begin{subequations}
\begin{align}
        \Pi_{2,\bk}^0(\bq, \Omega) &= \frac{1}{\beta}\sum_{\omega}\frac{-1}{(i\Omega+i\omega-\xi_{\bk+\bq})(i\omega-\xi_{\bk})} = \frac{\bv_\bk\cdot\bq}{\bv_\bk\cdot\bq-i\Omega}\delta(\xi_\bk), \\
\Pi_{2,\bk}^0(\bq,\tau) &= \frac{1}{\beta}\sum_{\Omega} e^{-i\Omega\tau} \Pi_{2,\bk}^0(\bq,\Omega)
= \abs{\bv_\bk \cdot \bq}e^{- \bv_\bk \cdot\bq \tau}\theta(\bv_\bk \cdot \bq \tau)\delta(\xi_\bk)
\end{align}   
\end{subequations}
From the above expression, it can be seen that $\Pi_{2,\bk}^0(\bq, \tau)$ decays on a time scale $|\tau| \sim \abs{\bv_\bk\cdot \bq}^{-1}$. 
Notice that in the Dyson equation (Eq.~\eqref{supp_eq: Dyson equation in time domain}) $\Pi^0_{2,\bk}$ is summed on $\bk$ over the entire Fermi surface, where the time scale for the decay of $\Pi^0_{2,\bk}$ is of order $(v_F |\bq|)^{-1}$ and much smaller than $(v_F\abs{q_{\perp}})^{-1}$ almost everywhere in the integral, except for a small region where $\bv_\bk \cdot \bq \sim v_F q_\perp$. We thus conclude that $\Lambda_\bk(\bq,\tau)$ is decaying at the time scale of $(v_F \abs{\bq})^{-1}$. On the other hand, as explained in the main text (and demonstrated explicitly in the following calculation),  $\Pi_{3,\bk}^0(\{\bq_a, \tau_a\})$ varies more slowly at the time scale of $(v_F\abs{q_{\perp}})^{-1}$ in the LWC limit where $\abs{q_{\perp}} \ll \abs{\bq_a}$. Therefore, in the LWC limit the time-integral in Eq.~\eqref{supp_eq: Pi3 with vertex correction, time domain} effectively evaluates the zero-frequency Fourier component of $\Lambda_\bk(\bq,\tau)$, which we denote simply as $ \Lambda_\bk \equiv \int d\tau \Lambda_{\bk}(\bq\rightarrow0,\tau)$. Correspondingly,  Eq.~\eqref{supp_eq: Dyson equation in time domain} becomes
\begin{equation}
    \Lambda_\bk = 1-\frac{1}{(2\pi)^2}\int _{S_F} d\bk'\Lambda_{\bk'} \frac{V_{\bk'\bk}}{\abs{\bv_{\bk'}}},
\end{equation}
which is Eq.~\eqref{eq: Dyson} in the main text. In the equal-time limit Eq.~\eqref{supp_eq: Pi3 with vertex correction, time domain} becomes
\begin{equation}
    s_3(\{\bq_a\}) =\int\frac{d^2\bk}{(2\pi)^2}\Lambda_\bk^3\; \Pi^0_{3,\bk}(\{\bq_a, 0\}).
\end{equation}
Now we turn to the detailed computation of $\Pi^0_{3,\bk}$.

\subsection{Computation of $\Pi^0_{3,\bk}$}
Recall the expression for the bare three-point bubble diagram:
\begin{equation}
\begin{split}
    \Pi^0_{3,\bk}(\{\bq_a, \tau_a\}) =
    \frac{1}{\beta^3}\sum_{\omega, \Omega_{2,3}} \frac{ e^{-i\Omega_2\tau_{21}-i\Omega_{3}\tau_{31}}}{(i\omega-\xi_\bk)(i\omega+i\Omega_2-\xi_{\bk+\bq_2})(i\omega+i\Omega_{23}-\xi_{\bk+\bq_{23}})} + \left(2\leftrightarrow 3\right).
\end{split}
\end{equation}
Let us first perform the $\omega$-sum, which can be achieved by noticing
\begin{equation}
\begin{split}
    \frac{1}{(i\omega-\xi_\bk)(i\omega+i\Omega_2-\xi_{\bk+\bq_2})(i\omega+i\Omega_{23}-\xi_{\bk+\bq_{23}})} =&
    \frac{\left(\frac{1}{i\omega-\xi_\bk}-\frac{1}{i\omega+i\Omega_2-\xi_{\bk+\bq_2}}\right)}{(i\Omega_{23}+\xi_\bk-\xi_{\bk+\bq_{23}})(i\Omega_2+\xi_\bk-\xi_{\bk+\bq_2})}\\
    -&\frac{\left(\frac{1}{i\omega+i\Omega_2-\xi_{\bk+\bq_2}}-\frac{1}{i\omega+i\Omega_{23}-\xi_{\bk+\bq_{23}}}\right)}{(i\Omega_{23}+\xi_\bk-\xi_{\bk+\bq_{23}})(i\Omega_3+\xi_{\bk+\bq_2}-\xi_{\bk+\bq_{23}})},
\end{split}
\end{equation}
and hence (with $f_\bk = \theta(E_F-E_\bk)$ the zero-temperature Fermi distribution)
\begin{equation}
\begin{split}
    \Pi^0_{3,\bk}(\{\bq_a, \tau_a\}) &=
    \frac{1}{\beta^2}\sum_{\Omega_{2,3}} e^{-i\Omega_2\tau_{21}-i\Omega_3\tau_{31}} \Big[ \frac{\left(f_\bk-f_{\bk+\bq_2}\right)}{(i\Omega_{23}+\xi_\bk-\xi_{\bk+\bq_{23}})(i\Omega_2+\xi_\bk-\xi_{\bk+\bq_2})} \\
    &\quad\quad\quad\quad\quad\quad\quad\quad\quad- \frac{\left(f_{\bk+\bq_2}-f_{\bk+\bq_{23}}\right)}{(i\Omega_{23}+\xi_\bk-\xi_{\bk+\bq_{23}})(i\Omega_3+\xi_{\bk+\bq_2}-\xi_{\bk+\bq_{23}})} \Big]+ \left(2\leftrightarrow 3\right) \\
    &=
    \frac{1}{\beta^2}\sum_{\Omega_{2,3}} e^{-i\Omega_{23}\tau_{31}-i\Omega_2\tau_{23}} \Big[ 
    \frac{\left(f_{\bk+\bq_{23}}-f_{\bk+\bq_3}\right)}{(i\Omega_{23}+\xi_\bk-\xi_{\bk+\bq_{23}})(i\Omega_2+\xi_{\bk+\bq_3}-\xi_{\bk+\bq_{23}})} \\
    &\quad\quad\quad\quad\quad\quad\quad\quad\quad-\frac{\left(f_{\bk+\bq_2}-f_\bk\right)}{(i\Omega_{23}+\xi_\bk-\xi_{\bk+\bq_{23}})(i\Omega_2+\xi_\bk-\xi_{\bk+\bq_2})}  \Big]+ \left(2\leftrightarrow 3\right).
\end{split}
\end{equation}
In the last equality, we have used $\Omega_2\tau_{21}+\Omega_3\tau_{31} = \Omega_{23}\tau_{31}+\Omega_2\tau_{23}$ and exchanged $2 \leftrightarrow 3$ for the second term inside the square bracket from the first equality. With this, we can organize the above expression compactly as
\begin{equation}
    \Pi^0_{3,\bk}(\{\bq_a, \tau_a\}) = 
    \frac{1}{\beta^2}\sum_{\Omega_{2,3}}\frac{e^{-i\Omega_{23}\tau_{31}}}{i\Omega_{23}+\xi_\bk-\xi_{\bk+\bq_{23}}}\Delta^{\bk+\bq_3}_\bk\left[\frac{e^{-i\Omega_2\tau_{23}}(f_{\bk+\bq_2}-f_\bk)}{i\Omega_2+\xi_\bk-\xi_{\bk+\bq_2}}\right] + (2\leftrightarrow 3),
\end{equation}
where $\Delta^{\bk+\bq}_\bk F(\bk) \equiv F(\bk+\bq) - F(\bk)$ defines the discrete derivative. To the second order in $\bq$, we have
\begin{equation}
    \Pi^0_{3,\bk}(\{\bq_a, \tau_a\}) = 
    \frac{1}{\beta^2}\sum_{\Omega_{2,3}}\frac{e^{-i\Omega_{23}\tau_{31}}}{i\Omega_{23}-\bv_\bk\cdot\bq_{23}}\bq_3\cdot\nabla_\bk\left[\frac{e^{-i\Omega_2\tau_{23}}}{i\Omega_2-\bv_\bk\cdot\bq_2} \bq_2\cdot\nabla_\bk f\right] + (2\leftrightarrow 3).
\end{equation}
Carrying out the bosonic Matsubara summations using $\frac{1}{\beta}\sum_\Omega \frac{e^{-i\Omega\tau}}{i\Omega-\bv_\bk\cdot \bq} \overset{\beta\rightarrow\infty}{=} -\text{sgn}(\tau)\theta(\bv_\bk\cdot\bq\tau)e^{-\bv_\bk\cdot\bq\tau}$, we obtain (defining $s_{ab}\equiv \text{sgn}(\tau_{ab})$)
\begin{equation}
\begin{split}
    \Pi^0_{3,\bk}(\{\bq_a, \tau_a\}) = \;\;&e^{-\bv_\bk\cdot\bq_{23}\tau_{31}} s_{31}\theta(s_{31}\bv_\bk\cdot\bq_{23})\bq_3\cdot\nabla_\bk[e^{-\bv_\bk\cdot\bq_2\tau_{23}}s_{23}\theta(s_{23}\bv_\bk\cdot\bq_2) \bq_2\cdot\nabla_\bk f]+(2\leftrightarrow 3).
\end{split}
\end{equation}
As far as the singular terms (i.e., those that contribute to $\abs{\bq_1\times\bq_2}$-singularity) are concerned, we could make use of the product rule and instead apply the derivative $\bq_3\cdot \nabla_\bk$ of the first term onto $\theta(s_{31}\bv_\bk\cdot \bq_{23})$ (and similarly for the second term with $\bq_2\cdot \nabla_\bk$ acting on $\theta(s_{21}\bv_\bk\cdot \bq_{23})$). In doing so, we have only dropped terms that are regular in $\bq$'s. We can thus recast the above expression as \footnote{From now on, whenever we refer to $\Pi^0_{3,\bk}$, we only mean its singular part that contributes to the $\abs{\bq_1\times\bq_2}$-singularity.}
\begin{equation}\label{supp_eq: Pi0_3k}
\begin{split}
    \Pi^0_{3,\bk}(\{\bq_a, \tau_a\}) &= e^{-\bv_\bk\cdot\sum_{a=1}^3\bq_a\tau_a} s_{32} \theta(s_{32}\bv_\bk\cdot\bq_3)[\bq_3\cdot\nabla_\bk\theta(\bv_\bk\cdot\bq_{23}) \bq_2\cdot\nabla_\bk f-(2\leftrightarrow 3)] \\
    &= e^{-\bv_\bk\cdot\sum_{a=1}^3\bq_a\tau_a} s_{32} \theta(s_{32}\bv_\bk\cdot\bq_3)[\bq_2\cdot\nabla_\bk\theta(\bv_\bk\cdot\bq_{1}) \bq_3\cdot\nabla_\bk f-(2\leftrightarrow 3)].
\end{split}
\end{equation}
In the above we have also used $\bq_{23}=-\bq_1$. This is Eq.~\eqref{eq: Pi0_3k} in the main text.\\

To see the above contributes to the $\abs{\bq_1\times\bq_2}$-singularity, 
notice that the factor in the square bracket in Eq.~\eqref{supp_eq: Pi0_3k} can be expressed as
\begin{equation}
\begin{split}
&\;\;\;\;\left[\bq_3\cdot\nabla_\bk(\bv_\bk\cdot\bq_1) \bq_2\cdot\nabla_\bk\xi_\bk -\bq_2\cdot\nabla_\bk(\bv_\bk\cdot\bq_1) \bq_3\cdot\nabla_\bk\xi_\bk \right]\delta(\bv_\bk\cdot\bq_1)\delta(\xi_\bk)\\
&=\left\{(\bq_2\cdot\bv_\bk)[ \bq_3\cdot\nabla_\bk(\bv_\bk\cdot\bq_1)] - (\bq_3\cdot\bv_\bk)[ \bq_2\cdot\nabla_\bk(\bv_\bk\cdot\bq_1)] \right\}\delta(\bv_\bk\cdot\bq_1)\delta(\xi_\bk)\\
&=(\bq_2\times\bq_3)\cdot[\bv_\bk \times \nabla_\bk(\bv_\bk\cdot \bq_1)]\delta(\bv_\bk\cdot\bq_1)\delta(\xi_\bk)\\
&\equiv \hat{z}\cdot(\bq_2\times \bq_3) J_\bk \delta(X_1)\delta(X_2)
\end{split}
\end{equation}
where in the last expression we have considered a change of variables from $(k_x,k_y) \rightarrow (X_1,X_2) \equiv(\xi_\bk,\bv_\bk\cdot\bq_1)$ which is associated to a Jacobian of the form
\begin{equation}
    J_\bk = \frac{\partial(X_1,X_2)}{\partial(k_x,k_y)} = \hat{z}\cdot\left[\bv_\bk \times \nabla_\bk(\bv_\bk\cdot\bq_1)\right].
\end{equation}
The delta functions indicate that $\Pi^0_{3,\bk}$ is only contributed by the Fermi surface critical points ($\bk_p$) where $\bv_p \cdot \bq_1 = 0$. Hence, 
\begin{equation}\label{supp_eq: intermediate result for Pi0_3k}
    \Pi^0_{3,\bk}(\{\bq_a, \tau_a\}) = \sum_p e^{-\bv_p\cdot \sum_{a=1}^3 \bq_a\tau_a} s_{32}\theta(s_{32}\bv_p\cdot\bq_3) [\hat{z}\cdot(\bq_2\times \bq_3)] \sgn(J_{\bk_p}) \delta^2(\bk-\bk_p).
\end{equation}
To further simplify the above expression, notice that $\bq_1 = \text{sgn}[(\bv_p\times \bq_1)\cdot\hat{z}](\hat{z}\times \bv_p)$, thus
\begin{equation}
\begin{split}
    J_{\bk_p} &= \text{sgn}[(\bv_p\times \bq_1)\cdot\hat{z}] \hat{z}\cdot\left(\bv_\bk \times \nabla_\bk(\bv_\bk\cdot(\hat{z}\times \bv_p))\right)\vert_{\bk=\bk_p} \\
    &=\text{sgn}[(\bv_p\times \bq_1)\cdot\hat{z}] \hat{z}\cdot\left(\bv_\bk \times \nabla_\bk\right)(\hat{z}\cdot(\bv_p\times \bv_\bk))\vert_{\bk=\bk_p}\\
    &=\text{sgn}[(\bv_p\times \bq_1)\cdot\hat{z}]\hat{z}\cdot\left(\bv_\bk \times [\hat{z}\cdot(\bv_\bk\times\nabla_\bk)\bv_\bk]\right)\vert_{\bk=\bk_p},
\end{split}
\end{equation}
implying
\begin{equation}
   \text{sgn}(J_{\bk_p}) = \text{sgn}[(\bv_p\times \bq_1)\cdot\hat{z}]\eta_p\;, \quad \text{with}\quad \eta_p \equiv \text{sgn}[\hat{z}\cdot\left(\bv_\bk \times [\hat{z}\cdot(\bv_\bk\times\nabla_\bk)]\bv_\bk\right)]\vert_{\bk=\bk_p}.
\end{equation}
We have introduced here the signature of the critical point $\eta_p$,
which equals to $\pm 1$ for a convex/concave critical point (i.e., with a neighborhood described by an electron/hole-like Fermi surface). This can be appreciated by considering an example where $\bv_p \parallel +\hat{x}$, then $\eta_p = \text{sgn}(\frac{\partial v_{\bk,y}}{\partial k_y})\vert_{\bk=\bk_p}$ which clearly reflects the convex/concave nature of the critical point. \\

Finally, substituting back into Eq.~\eqref{supp_eq: intermediate result for Pi0_3k} and noticing that $s_{32}$ can be replaced by $\text{sgn}(\bv_p\cdot \bq_3)$ due to the $\theta$-function, with $\text{sgn}(\bv_p\cdot\bq_3) \text{sgn}[(\bv_p\times\bq_1)\cdot\hat{z}] = \text{sgn}[\bq_3\cdot(\bq_1\times \hat{z})] = \text{sgn}[\hat{z}\cdot(\bq_2\times \bq_3)]$, we obtain
\begin{equation}
\begin{split}
    \Pi^0_{3,\bk}(\{\bq_a, \tau_a\})
    &=\abs{\bq_1\times\bq_2}\sum_p\eta_p \delta^2(\bk-\bk_p) e^{-\bv_p\cdot\sum_{a=1}^3\bq_a\tau_a}\theta(\bv_p\cdot\sum_{a=1}^3\bq_a\tau_a).
\end{split}
\end{equation}
This result corresponds to Eq.~\eqref{eq: Pi0_3k simplified} in the main text. In the above we have used $\bv_p\cdot\bq_1=0$, and hence $\theta(s_{32}\bv_p\cdot\bq_3) = \theta(\tau_{32}\bv_p\cdot\bq_3) = \theta(\bv_p\cdot\bq_3\tau_3 +\bv_p\cdot\bq_2\tau_2+\bv_p\cdot\bq_1\tau_1)$. Notice that $e^{-\bv_p\cdot\sum_{a=1}^3\bq_a\tau_a} = e^{-\bv_p\cdot\bq_2\tau_{23}}$, thus we have explicitly demonstrated that the singular part of $\Pi^0_{3,\bk}(\{\bq_a, \tau_a\})$ indeed varies at the time scale of $\abs{\bv_p\cdot\bq_2}^{-1} = (v_F\abs{q_\perp})^{-1}$.

\section{Three-body interaction}\label{supp_sec: 3-bdy interaction}
\setcounter{equation}{0}
\setcounter{figure}{0} 

\begin{figure}[H]
    \centering
    \resizebox{\columnwidth}{!}{\includegraphics[]{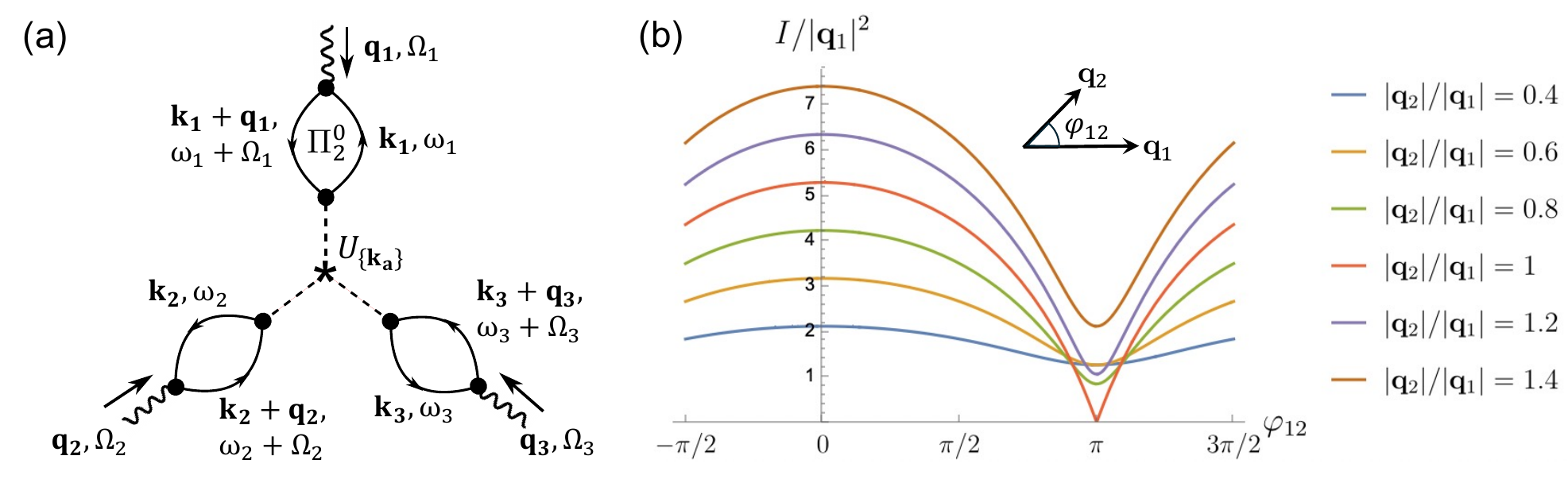}}
    \caption{(a) Feynman diagram for the contribution to the three-point density correlation from a three-body density-density interaction $U_{\{\bk_a\}}$. (b) Dependence of $I$, defined in Eq.~\eqref{supp_eq: delta s_3 from 3-body}, on the angle $\varphi_{12}$ between $\bq_1$ and $\bq_2$, showing that the three-body interaction does not generate $\abs{\bq_1\times\bq_2}$-type correction to $s_3$.}
    \label{supp_fig: Fig2}
\end{figure}

In this section we demonstrate that the three-body density-density interaction, 
$H_\text{3-bdy}=- \Pi_i\int \frac{d^2\bk_i}{(2\pi)^2}U_{\{\bk_{1,2,3}\}}\int\frac{d^2\bq_{1}d^2\bq_2}{(2\pi)^4}
c^\dagger_{\bk_3-\bq_1-\bq_2} c_{\bk_3} c^\dagger_{\bk_2+\bq_2}c_{\bk_2}c^\dagger_{\bk_1+\bq_1}c_{\bk_1}$, with all momenta restricted to a thin shell surrounding the Fermi surface, which is irrelevant (in the renormalization group sense) in the Fermi liquid theory, does not lead to any singular $\abs{\bq_1\times \bq_2}$-type correction to the equal-time three-point density correlation $s_3$. Following the diagram in Fig.~\ref{supp_fig: Fig2}(a), the corresponding correction to $s_3$ is
\begin{equation}
    \delta s_3(\{{\bf q}_a\}) =  \int \frac{d^2\bk_1d^2\bk_2d^2\bk_3}{(2\pi)^6}\frac{1}{\beta^2}\sum_{\Omega_{1,2}} U_{\{\bk_a\}}\prod_{a=1}^3\Pi^0_2({\bf k}_a,{\bf q}_a,i\Omega_a),
\end{equation}
with $\bq_3 = -\bq_1-\bq_2$, $\Omega_3=-\Omega_1-\Omega_2$, and the (momentum-resolved) polarization bubble is
\begin{equation}
    \Pi_2^0({\bf k},{\bf q},i\Omega) = \frac{1}{\beta}\sum_\omega \frac{-1}{(i\omega-\xi_\bk)(i\omega+i\Omega-\xi_{\bk+\bq})} = \frac{{\bf v}_{\bf k}\cdot {\bf q}}{{\bf v}_{\bf k}\cdot {\bf q}-i\Omega}\delta(\xi_{\bf k}),
\end{equation}
with $\xi_\bk \equiv E_\bk-E_F$ and we have taken the zero-temperature and small-$\bq$ limit. 
We thus have to compute
\begin{equation}\begin{split}
    &\int\frac{d\Omega_1 d\Omega_2}{(2\pi)^2} \frac{1}{i\Omega_1 - {\bf v}_1\cdot{\bf q}_1}\frac{1}{i\Omega_2 - {\bf v}_2\cdot{\bf q}_2}
    \frac{1}{i\Omega_1+i\Omega_2 + {\bf v}_3\cdot{\bf q}_3}\\
    &  = \frac{ (2\pi i)^2}{(2\pi)^2} \frac{[\theta(\bv_3\cdot\bq_3)-\theta(-\bv_1\cdot\bq_1)][\theta(-\bv_2\cdot\bq_2)-\theta(\bv_1\cdot\bq_1+\bv_3\cdot\bq_3)]}
    { {\bf v}_1 \cdot {\bf q}_1 + {\bf v}_2 \cdot {\bf q}_2+ {\bf v}_3 \cdot {\bf q}_3 } \\
   &= \frac{\theta_1 \theta_2 \theta_3 + \bar\theta_1 \bar\theta_2 \bar\theta_3}  { {\bf v}_1 \cdot {\bf q}_1 + {\bf v}_2 \cdot {\bf q}_2+ {\bf v}_3 \cdot {\bf q}_3 }
\end{split}\end{equation}
where we have defined ${\bf v}_a \equiv {\bf v}_{{\bf k}_a}$, $\theta_a \equiv \theta({\bf v}_a \cdot {\bf q}_a)$, and $\bar\theta_a \equiv 1-\theta_a$. When going from the second line to the third line, we also used the identity $(\theta_3-\bar{\theta}_1)\theta(\bv_1\cdot\bq_1+\bv_3\cdot\bq_3) = \theta_1\theta_3$.
Thus,
\begin{equation}
    \delta s_3 =  \int \frac{d^2\bk_1d^2\bk_2d^2\bk_3}{(2\pi)^6} (\theta_1 \theta_2 \theta_3 + \bar\theta_1 \bar\theta_2 \bar\theta_3)
    \frac{({\bf v}_1\cdot {\bf q}_1)({\bf v}_2\cdot {\bf q}_2)( {\bf v}_3\cdot {\bf q}_3)} { {\bf v}_1 \cdot {\bf q}_1 + {\bf v}_2 \cdot {\bf q}_2+ {\bf v}_3 \cdot {\bf q}_3 } U_{\{\bk_a\}}\delta(\xi_{\bk_1})\delta(\xi_{\bk_2})\delta(\xi_{\bk_3}).
\end{equation}
Integrating over $\xi_{\bf k}$ to obtain integrals over the Fermi surface, we obtain
\begin{equation}
    \delta s_3 =  \int_{S_F} \frac{d\bk_1 d\bk_2 d\bk_3}{(2\pi)^6} 
     (\theta_1 \theta_2 \theta_3 + \bar\theta_1 \bar\theta_2 \bar\theta_3)
    \frac{( \hat{\bf v}_1\cdot {\bf q}_1)( \hat{\bf v}_2\cdot {\bf q}_2)(  \hat{\bf v}_3\cdot {\bf q}_3)} { {\bf v}_1 \cdot {\bf q}_1 + {\bf v}_2 \cdot {\bf q}_2+ {\bf v}_3 \cdot {\bf q}_3 }
     U_{\{\bk_a\}}
\end{equation}
Note that due to the $\theta$'s, the denominator will only vanish if ${\bf v}_a\cdot {\bf q}_a$ are all zero, but then the numerator also vanishes. Next, let us see that $\delta s_3$ has no singular behavior of the form $\abs{\bq_1\times \bq_2}$ in general.\\

To expose any potential non-analytic momentum dependences, it suffices to consider an isotropic Fermi surface ($\abs{\bv_\bk}=v_F$) with a constant 3-body interaction $U_{\{\bk\}}=U$ (since they are generally analytic in momenta). Then $\delta s_3(\{\bq_a\}) = \frac{2k_F^3U}{(2\pi)^6} I(\abs{\bq_1},\abs{\bq_2},\varphi_{12})$ with
\begin{equation}\label{supp_eq: delta s_3 from 3-body}
    I(\abs{\bq_1},\abs{\bq_2},\varphi_{12})=\abs{\bq_1}\abs{\bq_2}\abs{\bq_3}\int_{-\pi/2}^{\pi/2}d\phi_1d\phi_2d\phi_3\frac{\cos\phi_1 \cos\phi_2 \cos\phi_3}{\abs{\bq_1}\cos\phi_1+\abs{\bq_2}\cos\phi_2+\abs{\bq_3}\cos\phi_3},
\end{equation}
where $\varphi_{12}$ is the angle between $\bq_1$ and $\bq_2$,  $\abs{\bq_3} = \sqrt{\abs{\bq_1}^2+\abs{\bq_2}^2+2\abs{\bq_1}\abs{\bq_2}\cos\varphi_{12}}$, and $\phi_{a=1,2,3}$ is the angle between $\bk_a$ and $\bq_a$. Figure~\ref{supp_fig: Fig2}(b) shows the dependence of $I$ as a function of $\varphi_{12}$ for different ratios $\abs{\bq_2}/\abs{\bq_1}$, which is generically analytic, except for one special case when $\bq_1=-\bq_2$ (i.e., the red curve at $\varphi_{12}=\pi$). But notice that in this very special case, $\bq_3=0$, thus we are \textit{outside of} the LWC limit (c.f. Eq.~\eqref{eq: q-collinear limit}, or pictorially, with the momentum-space triangle formed by $\{\bq_{1,2,3}\}$ forming a skinny obtuse triangle). Figure~\ref{supp_fig: Fig2}(b) indicates that the 3-body interaction \textit{does not} generate in $s_3$ the type of $\abs{\bq_1\times \bq_2}$-singularity that we are focusing on in this work. More analytically, suppose $\abs{\bq_1}=\abs{\bq_2}=q$, and hence $\abs{\bq_3}=2q\abs{\cos\frac{\varphi_{12}}{2}}$, we have
\begin{equation}
 \delta s_3 =  \frac{4q^2 k_F^3 U}{(2\pi)^6v_F} \int_{-\pi/2}^{\pi/2} d\phi_1 d\phi_2 d\phi_3 
    \frac{2\abs{\cos\frac{\varphi_{12}}{2}}\cos\phi_1 \cos\phi_2 \cos\phi_3}{\cos\phi_1 + \cos\phi_2 + 2\abs{\cos\frac{\varphi_{12}}{2}} \cos\phi_3}
\end{equation}
We can expand for small $\varphi_{12} \approx 0$:
\begin{equation}\begin{split}
    \delta s_3 = &\frac{ 4q^2 k_F^3 U}{(2\pi)^6v_F} \int_{-\pi/2}^{\pi/2} d\phi_1 d\phi_2 d\phi_3 \Bigg[
    \frac{2\cos\phi_1 \cos\phi_2 \cos\phi_3}{\cos\phi_1 + \cos\phi_2 + 2\cos\phi_3 }
    -\frac{1}{4}\varphi_{12}^2  \frac{\cos\phi_1\cos\phi_2\cos\phi_3(\cos\phi_1+\cos\phi_2)}{(\cos\phi_1+\cos\phi_2+2\cos\phi_3)^2}\Bigg]
\end{split}\end{equation}
Both integrals are convergent.   While there is a finite contribution at order $q^2$, there is \textit{no singularity} when $\varphi_{12}=0$, hence no singularity of form $\abs{\bq_1\times\bq_2}$ in the LWC limit.\\

Lastly, let us comment on the long-range correlation contributed by the three-body interactions. First taking the limit $\abs{\bq_3} \rightarrow 0$, we have $I(\abs{\bq_1},\abs{\bq_2},\varphi_{12}) \propto \abs{\bq_1} \abs{\bq_3}$, whose Fourier transform gives $\delta\mathfrak{s}_3(\{\br_{ij}\}) \propto \int d^2\bq_1d^2\bq_3 \abs{\bq_1}\abs{\bq_3}e^{i\bq_1\cdot\br_{12}+i\bq_3\cdot\br_{32}} \propto \abs{\br_{12}}^{-3} \abs{\br_{32}}^{-3}$. Thus, while the three-body interaction contributes to a long-range correlation, it is not favoring a collinear configuration for $\br_{1,2,3}$ and indeed decays much faster than the collinear straight-line correlation associated to the $\abs{\bq_1\times\bq_2}$-singularity, which decays as $\abs{\br_{ij}}^{-3/2}$ (c.f. Eq.~\eqref{eq: general Fermi surface_r-space}). This is also consistent with the expectation for the effect of an irrelevant interaction. 

\end{document}